\documentclass{ws-pp}

\def\be{\begin{equation}}
\def\ee{\end{equation}}
\def\eq#1{(\ref{#1})}
\def\bgrk#1{\mbox{{\boldmath $#1$ \unboldmath}}\!\!}
\def\siml{\,\hbox{\kern.1em \lower.6ex \hbox{$\sim$} \kern-1.12em
          \raise.6ex \hbox{$<$} }}
\def\simg{\,\hbox{\kern.1em \lower.6ex \hbox{$\sim$} \kern-1.12em
          \raise.6ex \hbox{$>$} }}

\begin{document}

\markboth{Brack {\it et al.}}{Periodic orbit theory including 
                              spin degrees of freedom}

%
\catchline{}{}{}{}{}
%

\title{PERIODIC ORBIT THEORY INCLUDING SPIN DEGREES OF FREEDOM
\footnote{Work supported by Deutsche Forschungsgemeinschaft through 
Forschergruppe FOR370 ``Spin electronics'' and Graduiertenkolleg 638 
``Nonlinearity and nonequilibrium in condensed matter''}
}

\author{\footnotesize MATTHIAS BRACK, CHRISTIAN AMANN, MIKHAIL PLETYUKHOV 
                      and OLEG ZAITSEV}

\address{Institute of Theoretical Physics, University of Regensburg\\
D-93040 Regensburg, Germany\\
e-mail: matthias.brack@physik.uni-regensburg.de}

\maketitle

\begin{history}
\received{(24. September 2003)}
\revised{(revised date)}
\end{history}

\begin{abstract}
We summarize recent developments of the semiclassical description
of shell effects in finite fermion systems with explicit inclusion
of spin degrees of freedom, in particluar in the presence of
spin-orbit interactions. We present a new approach that makes use
of spin coherent states and a correspondingly enlarged classical
phase space. Taking suitable limits, we can recover some of the
earlier approaches. Applications to some model systems are presented.
\end{abstract}

\section{Introduction}

The periodic orbit theory (POT), initiated over 30 years ago by 
Gutzwiller,\cite{gutz} is a semiclassical approach in which the level 
density of a quantum system is approximated in terms of the periodic 
orbits of the corresponding classical system through the so-called
`trace formula'. It has provided a great stimulus to the research area of 
quantum chaos,\cite{gubook,pot92,chaos} but is also applicable to 
integrable and nearly integrable systems.\cite{bablo,strma,beta,crli} 
Although originally developed to describe the motion of a particle bound 
in a given external potential or an ideally reflecting boundary (a 
so-called `quantum billiard'), the POT can also be applied to describe 
the quantum oscillations (shell-corrections) in many-fermion systems 
within the mean-field approximation.\cite{strma,bbook} For a general 
introduction to the POT and its applications in nuclear, mesoscopic and 
nanostructure physics, we refer to a recent text book.\cite{bbook} 

\begin{figure}[th]
\centerline{\psfig{file=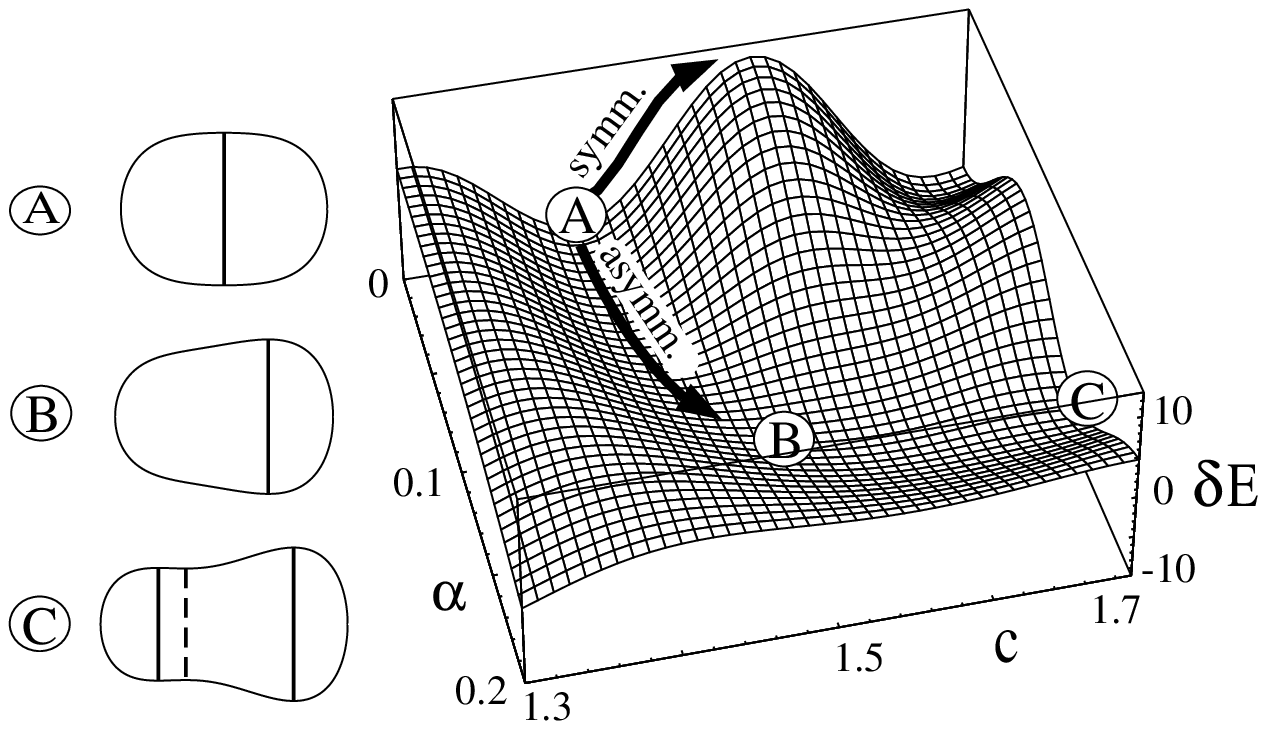,height=3.75cm}
\hspace*{0.05cm}\psfig{file=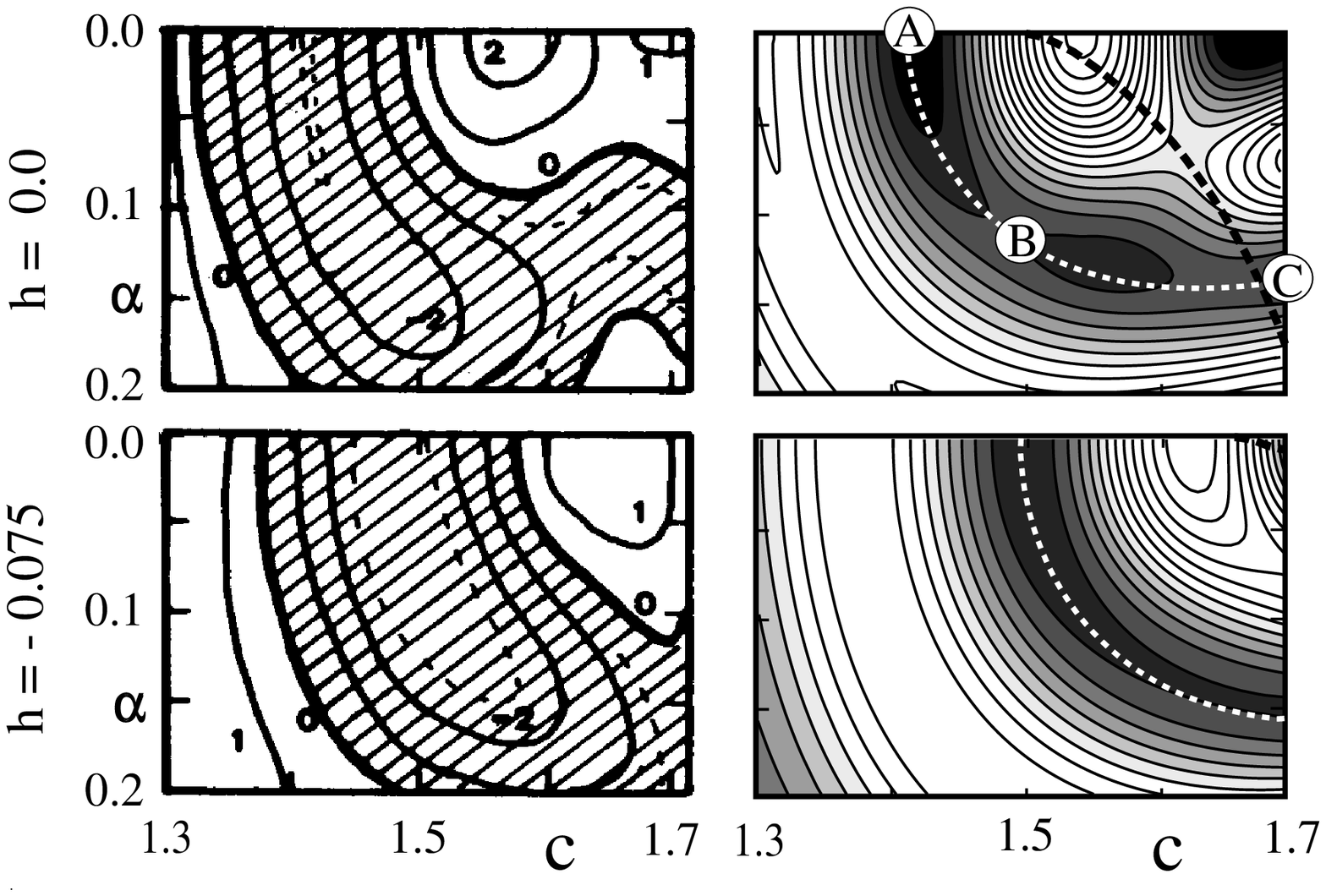,width=5.95cm}}
\caption{Shell-correction energy of $^{240}$Pu near fission isomer 
and outer barrier {\it vs.} elongation $c$ and mass asymmetry $\alpha$. 
{\it Left:} 3d plot of semiclassical POT calculation,$^{12}$ along 
with the shapes at isomer minimum (A), asymmetric saddle (B), and on the 
way towards scission (C) (perpendicular lines indicate planes containing 
the shortest periodic orbits).
{\it Right:} contour plot for neck parameter $h=0$ (upper panels) and
$h=-0.075$ (lower panels). The left panels show the quantum-mechanical
calculations with realistic Woods-Saxon potentials including pairing
correlations and spin-orbit interaction;$^{14}$ the right panels
show the semiclassical POT calculation for simple cavities.$^{12}$}
\label{fiss}
\end{figure}

An early application to nuclear physics consisted in a successful 
semiclassical explanation of the systematics of ground-state 
deformations.\cite{strdos} (A corresponding investigation was recently 
carried out also for metal clusters.\cite{pashk}) The onset of the mass 
asymmetry in the fission of actinide nuclei could also be explained
semiclassically.\cite{fiss,nobfis} Fig.\ \ref{fiss} shows the 
shell-correction energy around the outer fission barrier of $^{240}$Pu, 
plotted versus the elongation parameter $c$ and the mass asymmetry 
parameter $\alpha$. The 3d plot to the left and the contour plots to the 
extreme right were obtained within the POT in terms of the (few) shortest 
periodic orbits, modeling the nucleus by axially symmetric cavities (see 
shapes to the extreme left),\cite{fiss} whereas the contour plots next 
to the right are the old quantum-mechanical results using realistic 
deformed nuclear shell-model potentials.\cite{fuhil}

We can see that, in spite of the simplicity of the cavity model (with
only one kind of nucleons, neglecting pairing and spin-orbit interactions),
the semiclassical calculation reproduces almost quantitatively the correct
topology of the deformation energy surface and, in particular, predicts
correctly the mass-asymmetric adiabatic fission path. The latter, indicated
by the white dotted line, is simply given by the principle of stationary
action of the shortest periodic orbits. The wavefunctions of the
single-particle states, which are quantum-mechanically responsible for the 
lowering of the barrier due to the asymmetry, were found to have their 
maxima exactly in the planes containing the shortest periodic orbits that
dominate the asymmetry effect semiclassically.\cite{nobfis} 

\begin{figure}[th]
\centerline{\psfig{file=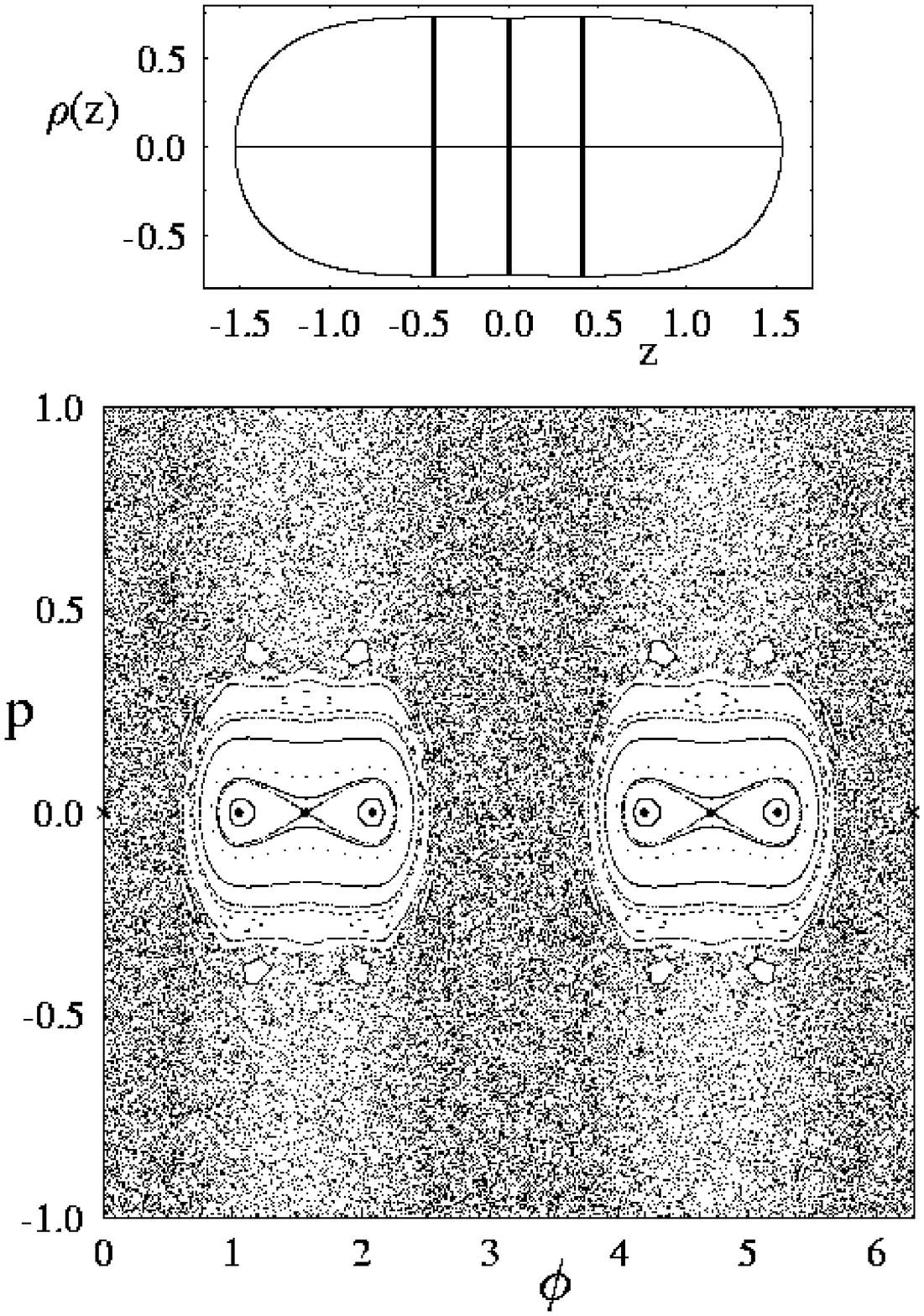,width=5.8cm}\hspace*{0.7cm}
\psfig{file=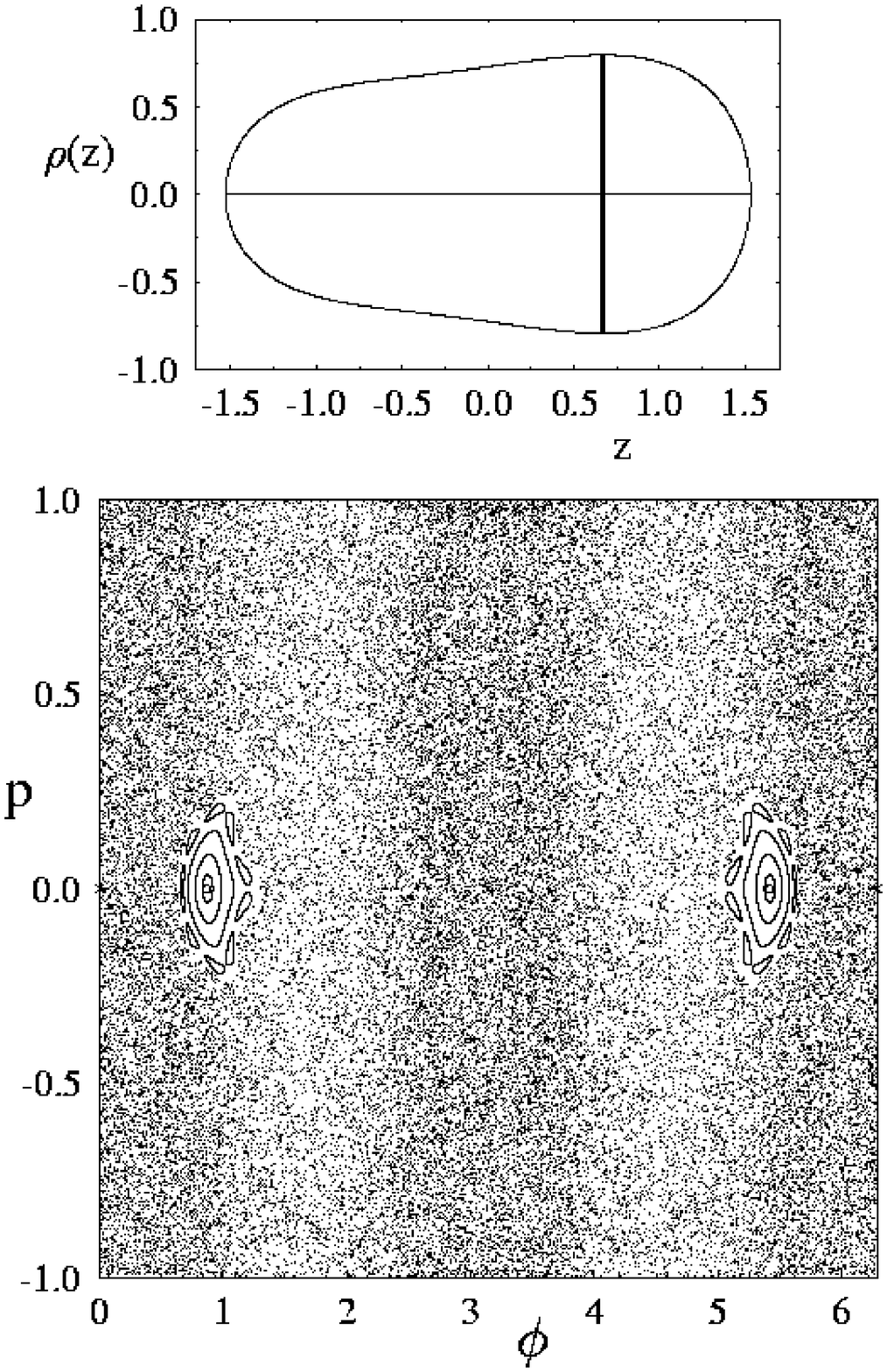,width=5.8cm}}
\vspace*{-0.1cm}
\caption{Poincar\'e surfaces of section $p$ (parallel momentum at
reflection point) {\it vs.} $\phi$ (polar angle of reflection point)
for angular momentum $L_z=0$ of the cavities corresponding to the 
symmetric outer barrier (left, $c=1.53$, $\alpha=0$) and to the 
asymmetric saddle (right, $c=1.53$, $\alpha=0.13$). The corresponding 
shapes and the planes containing the shortest orbits (vertical lines) 
are shown above the corresponding Poincar\'e plots.$^{13}$}
\label{poinc}
\end{figure}

That the classical motion of the nucleons at these deformations is almost 
chaotic can be seen in Fig.\ \ref{poinc}. Here we show Poincar\'e surfaces 
of section, obtained for angular momentum $L_z=0$, taken at the symmetric 
outer barrier (left) and at the asymmetric saddle (right); the corresponding
shapes with the planes containing the shortest periodic orbits are shown
on top of each plot. The regular motion is confined to small islands,
containing the relevant periodic orbits, surrounded by a chaotic sea. Note 
that the energetically favored asymmetric shape has much smaller regular 
islands than the symmetric shape. Nevertheless, the shell effect coming from 
these small regular regions of the classical dynamics is sufficiently strong 
to cause the collective asymmetry effect of the system.

This example illustrates the strength of the semiclassical theory in
explaining qualitatively (and, at times, even semi-quantitatively) 
important quantum effects, both in one-body and in many-body systems, 
in terms of classical dynamics. 

One aspect has, however, been neglected so far in the applications of the 
POT to nuclei: the spin of the nucleons. We know well that the spin-orbit 
interaction dramatically modifies the shell effects -- that is, after all, 
why it had to be introduced to make the nuclear shell model work in the 
first place.\cite{gmayer} In the semiclassical calculations of the 
Refs.\cite{strdos,fiss,nobfis} the neglect of the spin-orbit interaction 
was compensated by a simple readjustment of the Fermi energy, allowing one 
to locally reproduce the correct shell situations (or magic numbers). But 
this was only a temporary remedy, and a more rigorous semiclassical 
treatment of the spin-orbit interaction clearly remains highly desirable. 
The remainder of this paper is therefore devoted to some recent 
developments of the inclusion of spin degrees of freedom into the POT, 
which is by no means trivial since there is no classical analogon of the 
spin.

\section{Semiclassical theories with spin}

We limit the discussion here to systems of fermions with spin $s=1/2$
with a (mean-field) Hamiltonian linear in the spin operators
\be
\widehat{H} = \widehat{H}_0(\hat{\bf r},\hat{\bf p}) + \widehat{H}_{so}
\label{ham}
\ee
with
\be
\widehat{H}_0(\hat{\bf r},\hat{\bf p}) = \frac{\hat{\bf p}^2}{2m} + V({\bf r)}\,,
\qquad
\widehat{H}_{so} = \hbar\kappa\,{\bgrk{\sigma}}\cdot{\widehat{\bf C}}
                   (\hat{\bf r},\hat{\bf p})\,.
\label{ham01}
\ee
The second term is a general spin-orbit interaction, where $\widehat{\bf C}
(\hat{\bf r},\hat{\bf p})$ is an arbitrary vector function of coordinate 
$\hat{\bf r}$ and momentum operators $\hat{\bf p}$, and $\kappa$ is a coupling 
strength independent of $\hbar$. ${\bgrk{\sigma}}=(\sigma_x,\sigma_y,
\sigma_z)$ are the Pauli matrices defining the spin operators $\hat{\bf s}
=\frac12\hbar \bgrk{\sigma}$. In the non-relativistic reduction of the 
Dirac equation with an external electrostatic potential $V(\bf r)$, one
obtains
\be
\widehat{\bf C}(\hat{\bf r},\hat{\bf p}) = \left[\,\bgrk{\nabla} V({\bf r}) 
                                           \times\hat{\bf p}\,\right], \qquad 
                                           \kappa = 1/4m^2c^2\,.
\label{thomas}
\ee
With the Coulomb potential this yields the familiar Thomas term for
$\widehat{H}_{so}$.

\subsection{Earlier approaches}

We first summarize two earlier semiclassical approaches. 

($i$) {\bf SCL:} 
Littlejohn and Flynn\cite{lifl} developed a semiclassical theory for 
multi-com\-ponent systems, treating the spin matrices quantum mechanically 
while Wigner trans\-forming the matrix operator \eq{ham} to the classical 
phase space $({\bf r},{\bf p})$, keeping the leading terms in an $\hbar$ 
expansion. Diagonalisation leads to a pair of Hamiltonians
\be
H_\pm({\bf r,p}) = H_0({\bf r,p}) \pm \hbar\kappa |{\bf C}({\bf r,p})|\,,
                   \qquad (s=1/2)
\label{hpm}
\ee
where $H_0({\bf r,p})$ and ${\bf C}({\bf r,p})$ are the Wigner transforms
(to lowest order in $\hbar$) of the corresponding quantum operators. $H_\pm$ 
can be considered as two classical adiabatic Hamiltonians with opposite spin 
polarizations. Their two sets of periodic orbits must be superposed in the 
final trace formula. This approach is often referred to as the ``strong
coupling limit'' (SCL), since it becomes valid in the formal limit 
$\kappa\rightarrow\infty$ and $\hbar\rightarrow 0$ with $\hbar\kappa$ kept 
finite.\cite{lifl,boke} The SCL approach suffers, however, from the problem 
of {\it mode conversion} (MC): whenever $\bf C=0$ at a given point in (or in 
a subspace of) phase space, the two Hamiltonians $H_\pm$ become degenerate 
and singularities arise both in the classical equations of motion and in the 
calculation of the stabilities of the periodic orbits. (A similar situation 
occurs in the chemistry of molecular reactions when two or more adiabatic 
surfaces intersect.) The MC poses a difficult problem in semiclassical 
physics and chemistry that has not been satisfactorily solved so far. It 
was further discussed within the SCL in Refs.\cite{frgu,cham}

($ii$) {\bf WCL:}
Bolte and Keppeler\cite{boke} have derived a semiclassical theory from the 
Dirac equation. In the ``weak coupling limit'' (WCL) they arrive at a trace 
formula, in which the periodic orbits are given by the dynamics of the 
unperturbed Hamiltonian $H_0$ and the effect of spin precession around the 
local `magnetic field' $\kappa\,C({\bf r,p})$ appears through a simple
modulation factor. This approach neglects terms of higher than first order
in $\hbar\kappa$ and therefore is valid in the limit of weak spin-orbit
couplings. Furthermore, it yields only a trivial spin degeracy factor of 
two in the trace formula whenever all periodic orbits of $H_0$ are self
retracing ({\it i.e.}, librating between two turning points) such as in the
two systems illustrated below.

An instructive example is that of a two-dimensional electron gas with a 
Rashba type\cite{bych} spin-orbit interaction $\widehat{\bf C}=(-\hat{\pi}_y,
\hat{\pi}_x,0)$ in an external homogeneous magnetic field where 
$\hat{\bgrk{\pi}} = \hat{\bf p} - e {\bf A}/c$. For this system the exact 
quantum spectrum is explicitly known,\cite{bych} and analytical trace 
formulae have been given for both the exact quantum-mechanical level density 
and the semiclassical WC and SC limits,\cite{cham} from which the 
limitiations of these two approches become evident. 

In a successful application of the SCL, a model relevant for nuclear physics 
was investigated in Ref.\cite{cham} It consists of a three-dimensional 
harmonic oscillator with Thomas-type spin-orbit interaction \eq{thomas} 
\be
 V({\bf r}) = \sum_{i=x,y,z} \frac12\,\omega_i^2 r_i^2\,, \qquad  
 {\widehat H}_{so} = \hbar\kappa\,{\bgrk{\sigma}}\cdot
                     [\bgrk{\nabla}V({\bf r})\times\hat{\bf p}]\,, 
\label{v3ho}
\ee
which defines a Nilsson type\cite{nilss} Hamiltonian appropriate for light 
nuclei (where the $\ell^2$ term can be neglected). We express the oscillator 
frequencies in terms of two deformation parameters $\alpha$, $\beta$:
\be
\omega_x=\omega_0\,,\qquad \omega_y=(1+\alpha)\,\omega_0\,,\qquad
\omega_z=(1+\alpha)^\beta \omega_0\,,
\label{defpar}
\ee
and use $\hbar\omega_0$ as energy unit. For the general case of 
incommensurable frequencies ({\it i.e.}, three-axial deformations), all
periodic orbits of the unperturbed system $H_0$ are librations along the
coordinate axes and the WCL yields only the trivial spin factor of two.
The shortest periodic orbits of the Hamiltonians $H_\pm$ in the SCL lie
in the three coordinate planes and can be obtained analytically.\cite{cham}
For these orbits no mode conversion takes place, and the SCL can be used.

\begin{figure}[th]
\centerline{\psfig{file=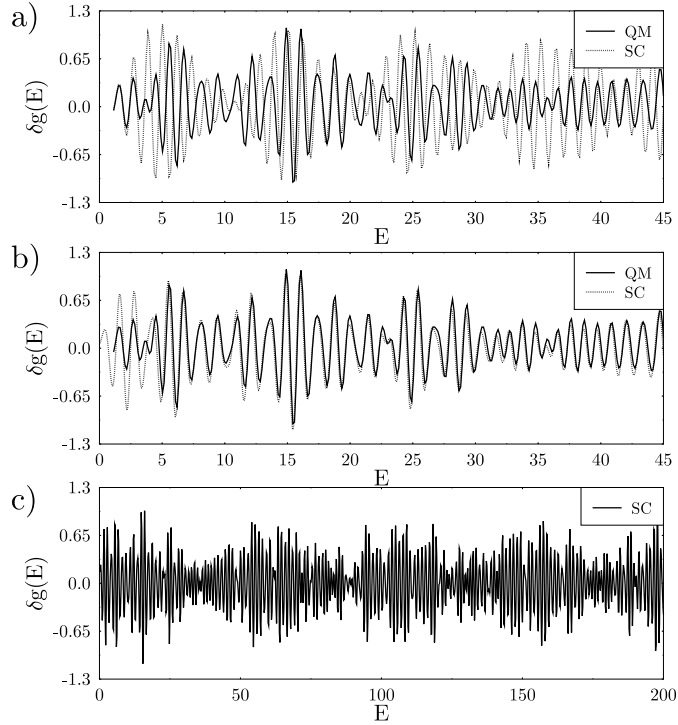,width=8.8cm}}
\vspace*{-0.3cm}
\caption{
Coarse-grained level density $\delta g(E)$ of the three-dimensional harmonic
oscillator \eq{v3ho} (see text for parameters, energy units: $\hbar\omega_0$). 
{\it Top and center:} solid lines give quantum-mechanical results (QM) with 
spin-orbit interaction. Dashed lines (SC) give semiclassical results (SC)
obtained a) without and b) with spin-orbit interaction. {\it Bottom:} c) 
SC as in b) over a larger energy region.$^{19}$ 
}
\label{3dho}
\end{figure}

In Fig.\ \ref{3dho} we show the shell correction to the level density of
this system, with a deformation $\alpha=0.1212$, $\beta=2$ and a spin-orbit 
strength $\kappa=0.1\,\omega_0^{-1}$. Both the quantum mechanical and the
semiclassical $\delta g(E)$ have been coarse grained by Gaussian convolution
over an energy range $\gamma=0.5\hbar\omega_0$, in order to suppress the
contribution of the longer orbits and hereby to emphasize the gross-shell
structure. The quantum-mechanical result is shown by the solid lines (QM) 
and includes the spin-orbit interaction in both curves a) and b). The 
semiclassical SCL result (SC) is shown by the dashed lines; in a) without 
spin-orbit interaction, which demonstrates that the latter dramatically 
changes the level density, and in b) with spin-orbit interaction. Only the 
six primitive planar orbits have been used. We see that this already leads 
to an excellent agreement with quantum mechanics, except at very low 
energies where semiclassics usually cannot be expected to work. As dicussed 
in Ref.\cite{cham}, bifurcations of the planar orbits occur for other 
deformations and values of $\kappa$. These can, in principle, be handled by 
suitable uniform approximations (see Ref.\cite{bbook} for an overview), but 
they complicate the semiclassical calculations numerically.

\subsection{POT in an extended phase space}

A new semiclassical approach has recently been presented\cite{plet,plza} in 
which the spin degrees of freedom were introduced through spin coherent 
states\cite{koch} $|z;s\rangle$ defined by
\be
|z;s\rangle = (1+|z|^2)^{-s}\exp(z\hat{s}_{+}/\hbar)\,|s,-s\rangle\,,\quad
\hat{s}_{-}|s,-s\rangle = 0\,, \quad \hat{s}_\pm = \hat{s}_x \pm i\hat{s}_y\,,
\ee
where $z=u-iv$ is a complex number. This allows one to define classical spin 
components ${\bf n}=(n_x,n_y,n_z)=\langle z;s|\hat{\bf s}|z;s\rangle/\hbar s$ 
and to enlarge the classical phase space by only {\it one} pair of canonical 
variables $(u,v)$, independently of the value of the spin $s$. Starting 
from the path integral in the 
SU(2) spin coherent state representation\cite{klakur} and making the usual
stationary-phase approximation in its evaluation, the semiclassical dynamics 
of the system in the {\it extended phase space} (EPS) $({\bf r,p},v,u)$ is 
then determined\cite{plet} by the Hamiltonian
\be
H({\bf r,p},v,u) = H_0({\bf r,p}) 
         + \hbar\kappa\,2s\,{\bf n}(v,u)\cdot{\bf C}({\bf r},{\bf p})\,.
\label{epsham}
\ee
Solving the equations of motion following from \eq{epsham}, one can 
determine the periodic orbits in the EPS and their properties, yielding the
required input into the Gutzwiller trace formula.\cite{gutz} Special
attention is required for the Maslov indices\cite{maslov} and other phases
arising in connection with the spin degrees of freedom.\cite{plza,phases} 

The EPS approach is free of the problem of mode conversion and, in principle,
applicable for both weak and strong spin-orbit interactions. Note that the
Hamiltonian \eq{epsham} explicitly couples the orbital degrees of freedom
$({\bf r,p})$ with the spin degrees of freedom $(u,v)$. This is illustrated
for the following model Hamiltonian of a two-dimensional semiconductor 
quantum dot with Rashba interaction:
\be
\widehat{H} = (\hat{p}_x^2 + \hat{p}_y^2)/2m^* 
            + m^*(\omega_x^2 x^2 + \omega_y^2 y^2)/2
            + \hbar\kappa\,(\sigma_y \hat{p}_x-\sigma_x\hat{p}_y)\,,
\label{hqdot}
\ee
where $m^*$ is the effective mass of the conduction electrons, and the
deformation parameters were chosen as $\omega_x=1.56\,\omega_0$ and 
$\omega_y=1.23\,\omega_0$. (In the figures, units are chosen such that 
$\hbar=m^*=\omega_0=1$ and E and $\kappa$ become dimensionless.) Like in 
the three-dimensional system discussed above, the periodic orbits of $H_0$ 
are pure librations, so that the WCL yields only trivial results. But here 
also the SCL cannot be used, due to the MC problem, so that a new treatment 
is required. 

For $0<\kappa\siml 0.7$, the following set of 12 shortest periodic orbits 
in the EPS were found:\cite{plet,plza} ($i$) Two pairs of orbits A$^\pm_x$ 
and A$^\pm_y$ librating along the $x$ and $y$ axes with fully polarized 
spin $n_y=\pm 1$ and $n_x=\pm 1$, respectively. ($ii$) Two pairs of orbits 
D$^\pm_{x1}$ and D$^\pm_{x2}$ oscillating around A$^\pm_x$ with $n_y\!\sim 
0$, and two pairs of orbits D$^\pm_{y1}$ and D$^\pm_{y2}$ oscillating 
around A$^\pm_y$ with $n_x\!\sim 0$. For stronger couplings with 
$\kappa\simg 0.7$, new orbits bifurcate from the A orbits.

\begin{figure}[th]
\vspace*{-1.0cm} 
\centerline{\psfig{file=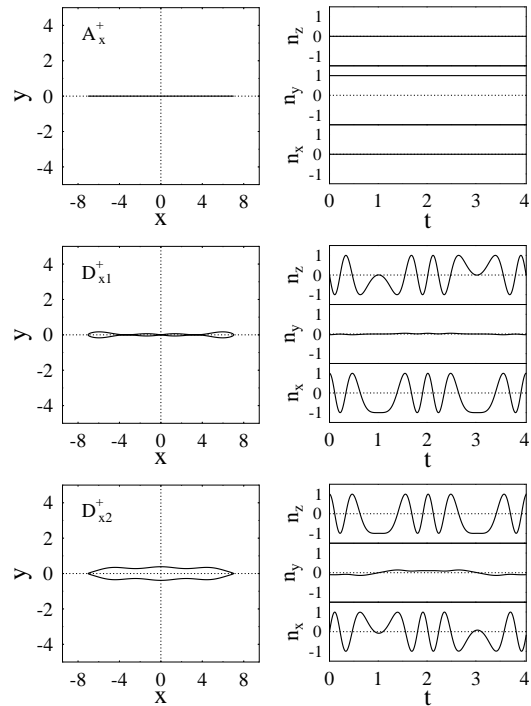,width=7.5cm}}
\vspace*{-0.3cm}
\caption{
Periodic orbits in the two-dimensional quantum dot \eq{hqdot} (see text 
for param\-eters). {\it Left panels:} orbits in the $(x,y)$ plane. {\it 
Right panels:} Spin components $n_x$, $n_y$, and $n_z$ versus time. {\it 
From top to bottom:} orbit A$^{\!+}_x$ along $x$ axis with polarized spin 
in $y$ direction, and orbits D$^+_{x1}$ and D$^+_{x2}$ oscillating around 
the $x$ axis with spin rotating near the $(n_x,n_z)$ plane.$^{22,23}$
}
\label{orbits}
\end{figure}

In Fig.\ \ref{orbits} we show the $(x,y)$ shapes of the orbits A$^{\!+}_{x}$, 
D$^+_{x1}$, and D$^+_{x2}$ (left panels), and the time dependence of their 
spin components $n_x$, $n_y$, and $n_z$ over one period (right panels), all 
evaluated for $\kappa=0.67$ and $E=60$. We see that along the orbits 
D$^+_{x1}$ and D$^+_{x2}$, the spin rotates mainly near the $(n_x,n_z)$ 
plane (i.e., $n_y\sim 0$), but in a non-uniform way. This complicated spin 
motion, together with the wiggly orbital shapes of the D orbits, reveals the 
rather sophisticated dynamics which is obtained from the equations of motion 
in the EPS through the explicit coupling of spin and orbital degrees of freedom.

\begin{figure}[th]
\vspace*{-1.0cm} 
\centerline{\psfig{file=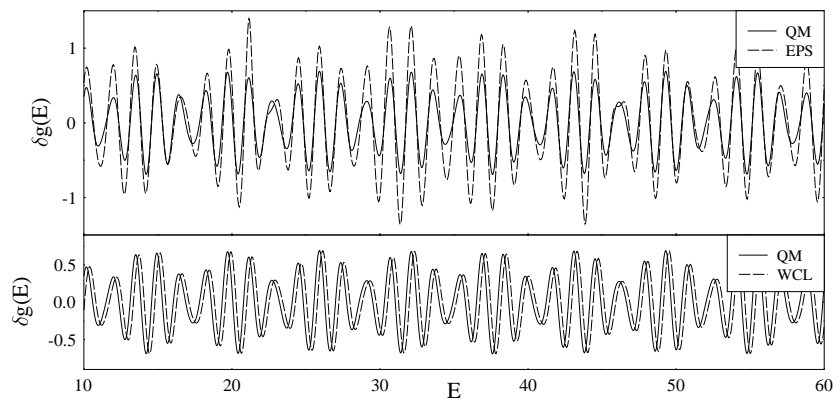,height=6.0cm}}
\vspace*{-0.5cm}
\caption{
Coarse-grained level density of quantum dot \eq{hqdot} for the same 
parameters as in Fig.\ \ref{orbits}. {\it Solid lines:} quantum-mechanical 
result. {\it Dashed lines:} semiclassical results; in upper panel: EPS
result, in lower panel: WCL result.$^{23}$
}
\label{levden}
\end{figure}

Fig.\ \ref{levden} shows the shell correction of the level density, obtained 
quantum-mech\-anically (solid lines) and semiclassically (dashed lines), both 
coarse-grained over an energy range $\gamma=0.6$. The upper panel contains the
semiclassical EPS result, while the lower panel exhibits the WCL result which
is identical to that obtained by ignoring the spin-orbit interaction and 
multiplying the level density by a spin factor of two. (Note that the 
semiclassical trace formula for the unperturbed harmonic oscillator is 
analytically known and quantum-mechanically exact.\cite{bbook}) We observe a 
reasonably good agreement of the EPS result with quantum mechanics. The 
semiclassical amplitudes are too large, which is attributed to close-lying 
bifurcations of the periodic orbits;\cite{plza} suitable uniform 
approximations are expected to remedy this defect. The phases of the 
quantum oscillations are, however, very well reproduced, which is not 
achieved at all in the WCL.

\newpage

\section{Summary and Outlook}

After a short review of the two earlier semiclassical approaches including
spin-orbit interactions, corresponding to the weak coupling limit (WCL) and 
the strong coupling limit (SCL), we have presented a new approach that makes
use of spin coherent states and leads to (semi)classical dynamics in an
extended phase space (EPS). Both the WCL and the SCL (with the `no-name'
phase still lacking) could be recovered from the EPS approach taking
suitable limits.

We have only discussed here Hamiltonians linear in spin and
systems with spin $s=1/2$. The EPS approach was formulated for arbitrary
spin-dependent Hamiltonians and arbitrary values of $s$. As is well known, 
semiclassical approximations work best in the limit of large quantum numbers.
Whether they can be used for small quantum numbers is a matter of numerical 
experience and fortune (like for harmonic oscillators without spin). As to 
the spin, there are firm grounds to expect good semiclassical results also 
for $s=1/2$ for Hamiltonians linear in spin.\cite{plza}

The example shown in Figs.\ \ref{orbits} and \ref{levden} suggests that the 
EPS approach has a good potential for a general semiclassical theory with spin. 
Practically, this approach suffers from a large number of bifurcations occurring 
for the periodic orbits under variations of both $\kappa$, energy, and deformation 
parameters. It would therefore be desirable to use it as a formal starting point 
for further approximations. 

Indeed, in the limit of a weak spin-orbit coupling such that $H_{so}\ll H_0$, the 
WCL trace formula of Ref.\cite{boke} could be rigorously derived from the EPS 
approach.\cite{oleg} Hereby the equations of motion of the orbital and spin 
degrees of freedom were decoupled, and the effects of ${\hat H}_{so}$ were 
included in the phases of the trace formula in first-order perturbation 
theory.\cite{crep} Pushing this treatment to second order, along the lines 
proposed in Ref.\cite{pert2} for spin-independent systems, might allow one to 
extend the WCL trace formula, so that it becomes valid for larger values of 
$\kappa$ while still benefitting from a simpler determination of the periodic 
orbits. In particular, for the self-retracing orbits where the first-order result 
is trivial, the second-order treatment is expected\cite{pert2} only to lead to a 
phase correction while the amplitudes in the trace formula still are determined 
by the unperturbed orbits of $H_0$. This expectation is, indeed, strongly
supported by the results shown in the lower part of Fig.\ \ref{levden}.

On the other hand, a careful study\cite{plza} of the situation $H_{so}\sim H_0$ 
reveals that the essential ingredients of the SCL approach of 
Refs.\cite{lifl,frgu} -- without, however, the so-called `no-name phase' -- 
can also be retrieved from the EPS approach. Adding suitable corrections to
this limit might help to overcome the mode conversion problem without going
through the cumbersome task of finding all relevant periodic orbits in the
extended phase space, including all possible bifurcations and their treatment 
by uniform approximations.

In a recent application of the EPS approach to mesoscopic transport theory, 
the effects of spin-orbit interaction on weak anti-localization have been 
investigated,\cite{zari} and its application to semiclassical studies of
nuclear shell structure including realistic shell-model potentials and
spin-orbit interactions is in progress.\cite{subra}

\end{document}